\documentstyle[prd,preprint,aps]{revtex}
\begin{document}
\draft

\title{Pair production and angular distribution of helicity flipped neutrinos 
       in a Left-Right symmetric model}

\author{A. Guti\'errez-Rodr\'{\i}guez $^1$, M. A. Hern\'andez-Ru\'{\i}z $^2$,
        M. Maya $^3$ and M. L. Ortega $^3$}

\address{(1) Facultad de F\'{\i}sica, Universidad Aut\'onoma de Zacatecas\\       
Apartado Postal C-580, 98060 Zacatecas, Zacatecas M\'exico.}

\address{(2) Facultad de Ciencias Qu\'{\i}micas, Universidad Aut\'onoma de Zacatecas\\
Apartado Postal 585 Zacatecas, Zacatecas M\'exico.}

\address{(3) Facultad de Ciencias F\'{\i}sico Matem\'aticas, Universidad Aut\'onoma de Puebla\\
Apartado Postal 1364, 72000 Puebla, Puebla M\'exico.}

\date{\today}
\maketitle

\begin{abstract}

We assume that a massive Dirac neutrino is characteized by two phenomenological
parameters, a magnetic moment, and a charge radius, and we calculate the
cross-section of the scattering $e^+e^-\to \nu \bar \nu$ in a left-right symmetric
model. We also analyze the angular distribution of the neutrino (antineutrino)
with respect to the original direction of the electron (positron) to different
state of helicity of the neutrino. We find that the favored directions for the
neutrino (antineutrino) with respect to the electron (positron) is forward
$(\theta =0)$ and backward $(\theta= \pi)$, and is not very probable in the
perpendicular direction $(\theta = \frac{\pi}{2})$. The calculation is for
$\phi = -0.005$ and $M_{Z_2}=500$ $GeV$, parameters of the Left-Right symmetric
model.

\end{abstract}


\narrowtext

\section{Introduction}
Of all the particles of the Standard Model (SM) \cite{S.L.Glashow}, neutrinos
are the least known. Because they are treated as massless particles, the
physical phenomena associated with them are rather limited. On the other hand,
in case of massive neutrinos, which are predicted by some Grand Unified
Theories \cite{Mohapatra}, several new effects can occur. Massive neutrinos
open up the possibility of a variety of new physical phenomena.

Neutrinos seem to be likely candidates for carrying features of physics beyond
the standard model. Not only masses and mixings, but also charge radius, magnetic moment
and electric dipole moment \cite{Degrassi,Lucio} are signs of new physics,
and are of relevance in terrestrial experiments, the solar neutrino problem
\cite{Voloshin,Vogel}, astrophysics and cosmology \cite{Cisneros,Salati}.

When the explosion of the Supernova 1987 A (SN 1987 A) occurred, 
the astrophysics of neutrinos was born. The observation of neutrinos from SN
1987 A \cite{Hirata,Bionta}, in fair agreement with predictions from supernova
models, has been used by several authors to bound the properties and interactions
of various exotic and non-exotic particles \cite{Barbieri}. The experimental
observation of SN 1987 A launched several new searches for supernova neutrinos.
Besides specially developed detectors, basically all new real-time solar neutrino
detectors like Super-Kamiokande, ICARUS and SNO will be able to see such neutrinos.

Thus we can conclude that, the study of the neutrino continues to be the subject
of current research, both theoretical and experimental.

At the present time, all the available experimental data for electroweak
processes can be well understood in the context of the Standard Model of the
electroweak interactions (SM) \cite{S.L.Glashow}, except the results of the
Super-Kamiokande experiment on the neutrino mass \cite{Fukuda}. Hence, the SM is
the starting  point of all the extended gauge models. In other words, any
gauge group with physical sense must have as a subgroup the
$SU(2)_{L} \times U(1)$ group of the standard model, in such a way that their
predictions agree with those of the SM at low energies. The purpose of the
extended theories is to explain some fundamental aspects which are not
clarified in the framework of the SM. One of these aspects is the origin of the
parity violation at the current energies. The Left-Right Symmetric Model (LRSM)
based on the $SU(2)_{R}\times SU(2)_{L}\times U(1)$ gauge group
\cite{J.C.Pati} gives an answer to that problem, since it restores the parity
symmetry at high energies and gives its violations at low energies as a
result of the breaking of gauge symmetry. Detailed discussions on LRSM can
be found in the literature \cite{J.C.Pati,R.N.Mohapatra,G.Senjanovic,G.Senjanovic1}.

Although in the framework of the SM, neutrinos are assumed to be electrically
neutral. Electromagnetic properties of the neutrino are discussed in many
gauge theories beyond the SM. Electromagnetic properties of the neutrino may
manifest themselves in a magnetic moment of the neutrino as well as in a
non-vanishing charge radius, both making the neutrino subject to the electromagnetic
interaction.

In this paper, we start from a Left-Right Symmetric Model (LRSM) with massive
Dirac neutrinos left and right-handed, with an electromagnetic structure that consists
of a charge radius $\langle r^2\rangle $ and of a anomalous magnetic moment $\mu_\nu$,
and we calculated the total cross-section of the scattering $e^+e^-\to \nu \bar \nu$.
We emphasize here the simultaneous contribution of the charge radius, of the
anomalous magnetic moment as well as of the additional $Z_{2}$ heavy gauge boson, and of the
mixing angle $\phi$ parameters of the LRSM to the cross-section.
The Feynman diagrams which contribute to the process $e^+e^-\to \nu \bar \nu$ are
shown in Fig. 1. We also analyzed the angular distribution of the neutrino (antineutrino)
with respect to the original direction of the electron (positron) to different
state of helicity of the neutrino. We find that the directions of the neutrino
(antineutrino) with respect to the electron (positron) is forward $(\theta =0)$
and backward $(\theta = \pi)$, and is not very probable in the perpendicular
direction $(\theta = \frac{\pi}{2})$.

This paper is organized as follows. In Sec. II we carry out the calculations
of the process $e^+e^-\to \nu \bar \nu$. In Sec. III we present the expressions for the
helicities. In Sec. IV we achieve the numerical computations. Finally, we summarize
our results in Sec. V.

\newpage

\section{The electron positron-neutrino antineutrino scattering}

In this section we obtain in the context of the LRSM the cross-section of the
process

\begin{equation}
e^-(p_1) + e^+(p_2)\to \bar \nu (k_1,\lambda_1)+\nu (k_2,\lambda_2),
\end{equation}

\noindent here $p_1$, $p_2$, $k_1$, and $k_2$ are the particle momenta and $\lambda_1$
($\lambda_2$) is the neutrino (antineutrino) helicity.

We will assume that a massive Dirac neutrino is characterized by two phenomenological
parameters, a magnetic moment $\mu_{\nu}$, expressed in units of the electron
Bohr magnetons, and a charge radius $\langle r^{2}\rangle$. Therefore, the expression
for the amplitude $\cal M$ of the process $e^-e^+ \to \nu \bar \nu$ Eq. (1)
due only to $\gamma$ and $Z^0$ exchange, according to the diagrams depicted in
Fig. 1 is given by

\begin{equation}
{\cal M}_\gamma =-ie^2 \bar \nu (k_2,\lambda_2)\frac{\Gamma ^\mu}{q^2}\nu (k_1,\lambda_1)
                    \bar e(p_2)\gamma _\mu e(p_1),
\end{equation}

\noindent with

\begin{equation}
\Gamma ^\mu = eF_1(q^2)\gamma ^\mu -\frac{ie}{2m_\nu}F_2(q^2)\sigma ^{\mu \nu}q_\nu,
\end{equation}

\noindent the neutrino electromagnetic vertex, where $q$ is the momentum
transfer and $F_{1,2}(q^2)$ are the electromagnetic form factors of the neutrino.
Explicitly \cite{Vogel}

\[
F_1(q^2)=\frac{1}{6}q^2 \langle r^2 \rangle, 
\]
\[
F_2(q^2)=-\mu _\nu \frac{m_\nu}{m_e}, 
\]

\noindent where, as already mentioned, $\langle r^2\rangle$ is the neutrino
mean-square charge radius and $\mu_{\nu}$ the anomalous magnetic moment. Therefore

\begin{equation}
{\cal M}_\gamma =-i\frac{e^2}{q^2}[F\bar \nu (k_2,\lambda_2)\gamma^\mu \nu (k_1,\lambda_1)
                    \bar e(p_2)\gamma _\mu e(p_1)
                 +GK^\mu\bar \nu (k_2,\lambda_2)\nu (k_1,\lambda_1)
                    \bar e(p_2)\gamma _\mu e(p_1)],
\end{equation}

\noindent with

\[
F=F_1+iF_2, \hspace*{5mm} G=-i\frac{F_2}{2m_\nu}, \hspace{5mm} ${\mbox and}$  \hspace*{5mm} K^\mu =(k_2-k_1)^\mu.
\]

\noindent Furthermore

\begin{eqnarray}
{\cal M}_{Z^0} &=&-i\frac{g^2}{8c^2_{W}(q^2-M^2_{Z_1})}[P\bar \nu (k_2,\lambda_2)\gamma^\mu \nu (k_1,\lambda_1)
                    \bar e(p_2)\gamma _\mu e(p_1)\nonumber\\        
             && +Q\bar \nu (k_2,\lambda_2)\gamma^\mu \gamma_5\nu (k_1,\lambda_1)
                    \bar e(p_2)\gamma _\mu e(p_1)\nonumber\\
             && +R\bar \nu (k_2,\lambda_2)\gamma^\mu \nu (k_1,\lambda_1)
                    \bar e(p_2)\gamma _\mu \gamma_5e(p_1)\\
             && +S\bar \nu (k_2,\lambda_2)\gamma^\mu \gamma_5\nu (k_1,\lambda_1)
                    \bar e(p_2)\gamma _\mu \gamma_5\gamma_5e(p_1)],\nonumber
\end{eqnarray}

\noindent where

\begin{eqnarray}
P&=&(A+2B+C)g_V,\nonumber\\
Q&=&(-A+C)g_A,\\
R&=&(-A+C)g_V,\nonumber\\
S&=&(A-2B+C)g_A\nonumber,
\end{eqnarray}

\noindent the constants $A$, $B$ and $C$ depend only on the LRSM, and are
given by \cite{Ceron}

\begin{eqnarray}
A&=&a^2+\Gamma c^2=(c_{\phi}-\frac{s^{2}_{W}}{r_{W}}s_{\phi})^{2}+\Gamma(\frac{{s^{2}_{W}}}{r_{W}}c_{\phi}+s_{\phi})^{2},\nonumber\\
B&=&ab+\Gamma cd=(c_{\phi}-\frac{{s^{2}_{W}}}{r_{W}}s_{\phi})(-\frac{{c^{2}_{W}}}{r_{W}}s_{\phi})
   +\Gamma(\frac{{s^{2}_{W}}}{r_{W}}c_{\phi}+s_{\phi})(\frac{{c^{2}_{W}}}{r_{W}}c_{\phi}),\nonumber\\
C&=&b^2+\Gamma d^2=(\frac{c^{2}_{W}}{r_{W}}s_{\phi})^{2}+\Gamma(\frac{c^{2}_{W}}{r_{W}}c_{\phi})^{2}, \nonumber
\end{eqnarray}

\noindent with

\[
\Gamma =\frac{q^2-M^2_{Z^0_1}}{q^2-M^2_{Z^0_2}}.
\]

\noindent  While $g_V=-\frac{1}{2}+2\sin^2\theta_W$ and $g_A=-\frac{1}{2}$,
according to the experimental data \cite{Review}.

The square of the amplitude is obtained by sum over spin states of the final fermions, so

\begin{equation}
\sum_{sp}|{\cal M}_{T}|^2=\sum_{sp}|{\cal M}_{\gamma}+{\cal M}_{Z^0}|^2
                          =\sum_{sp}(|{\cal M}_\gamma|^2
                          +|{\cal M}_{Z^0}|^2+{\cal M}_{Z^0 }{\cal M}^{\dagger}_\gamma
                          +{\cal M}^{\dagger}_{Z^0} {\cal M}_\gamma),
\end{equation}

\noindent where:

\newpage

\begin{eqnarray}
\sum_{sp}|{\cal M}_\gamma|^2&=&4H_1E^4
\{(F^2_1+F^2_2)(1+x^2)(1-\lambda_{\bar \nu}\lambda_\nu)\nonumber\\
&&-2F^2_2(x^2-1-\lambda_{\bar \nu}\lambda_\nu)+\frac{E^2}{m^2_\nu}F^2_2
(1-x^2)(1+\lambda_{\bar \nu}\lambda_\nu)\},\\
\sum_{sp}|{\cal M}_{Z^0}|^2&=&4H_2E^4
\{(P^2+Q^2+R^2+S^2)(1+x^2)(1-\lambda_{\bar \nu}\lambda_\nu)\nonumber\\
&&+4x(PS+QR)(1-\lambda_{\bar \nu}\lambda_\nu)
+2(PQ+RS)(1+x^2)(\lambda_\nu-\lambda_{\bar \nu})\nonumber\\
&&+4x(PR+QS)(\lambda_\nu-\lambda_{\bar \nu})\},\\
\sum_{sp}({\cal M}_{Z^0}{\cal M}^{\dagger}_{\gamma}
+{\cal M}^{\dagger}_{Z^0} {\cal M}_\gamma )&=& 8H_3E^4\{
F_1[P(1+x^2)(1-\lambda_{\bar \nu}\lambda_\nu)
+Q(1+x^2)(\lambda_\nu-\lambda_{\bar \nu})\nonumber\\
&&+2xR(\lambda_\nu-\lambda_{\bar \nu})
+2xS(1-\lambda_{\bar \nu}\lambda_\nu)]\nonumber\\
&&+F_2[P(2-\lambda_{\bar \nu}\lambda_\nu x^2)+Q(\lambda_\nu-\lambda_{\bar\nu})(1-\frac{1}{2}x^2)\nonumber\\
&&+\frac{3}{2}xR(\lambda_\nu-\lambda_{\bar\nu})+xS(2-\lambda_{\bar\nu}\lambda_\nu)]\},
\end{eqnarray}

\noindent with

\[
H_1=\frac{e^4}{q^4}, \hspace*{5mm} H_2=\frac{g^4}{64c^4_W(s-M^2_{Z^0_1})^2},
\hspace*{5mm} H_3=\frac{e^2g^2}{8c^2_Wq^2(q^2-M^2_{Z^0_1})},
\]
\noindent and $x=\cos\theta$, where $\theta$ is the scattering angle.

In the expressions (8), (9), and (10) the simultaneous contribution 
of the anomalous magnetic moment, of the charge radius electroweak, of the
heavy gauge boson $Z^0_R$ and of the mixing angle $\phi$ are observed.

The scattering cross-section in the center of mass system (where $s$ is the
square of the center-of-mass energy) is given by

\begin{equation}
\frac{d\sigma}{d\Omega}=\frac{1}{64\pi^2 s}\sum_{sp}|{\cal M}_T|^2,
\end{equation}

\noindent where the square of the total amplitude of transition ${\sum}_{sp}|{\cal M}_T|^2$
is given in the Eq. (7).

The differential cross-section to each contribution is

\begin{eqnarray}
(\frac{d\sigma}{d\Omega})_\gamma &=&\frac{\alpha^2}{16s}{\cal F}_1
(F_1,F_2,E,m_\nu,\lambda_{\bar \nu},\lambda_\nu,x),\\
(\frac{d\sigma}{d\Omega})_{Z^0}&=&\frac{\alpha^2}{64s}R^2_1(s){\cal F}_2
(P,Q,R,S,\lambda_{\bar \nu},\lambda_\nu,x),\\            
(\frac{d\sigma}{d\Omega})_{\gamma Z^0}&=&\frac{\alpha^2}{16s}R_1(s){\cal F}_3
(F_1,F_2,P,Q,R,S,\lambda_{\bar \nu},\lambda_\nu,x),
\end{eqnarray}

\noindent so that the total cross-section is

\begin{eqnarray}
(\frac{d\sigma}{d\Omega})_T &=&\frac{\alpha^2}{16s}{\cal F}_1
                               (F_1,F_2,E,m_\nu,\lambda_{\bar \nu},\lambda_\nu,x)\nonumber\\
                            &&+\frac{\alpha^2}{64s}R^2_1(s){\cal F}_2
                               (P,Q,R,S,\lambda_{\bar \nu},\lambda_\nu,x)\nonumber\\                   
                            &&+\frac{\alpha^2}{16s}R_1(s){\cal F}_3
                               (F_1,F_2,P,Q,R,S,\lambda_{\bar \nu},\lambda_\nu,x),
\end{eqnarray}

\noindent where

\begin{equation}
R_1(s)=\frac{s}{\sin^22\theta_W (s-M^2_{Z^0_1})},
\end{equation}

\noindent is the factor of resonance.

The kinematics to each contribution to be contained in the functions

\begin{eqnarray}
\gamma :{\cal F}_1&=&(F^2_1+F^2_2)(1+x^2)(1-\lambda_{\bar \nu}\lambda_\nu)
-2F^2_2(x^2-1-\lambda_{\bar \nu}\lambda_\nu)\nonumber\\
&&+\frac{E^2}{m^2_\nu}F^2_2(1-x^2)(1+\lambda_{\bar \nu}\lambda_\nu),\\
Z^0:{\cal F}_2&=&(P^2+Q^2+R^2+S^2)(1+x^2)(1-\lambda_{\bar \nu}\lambda_\nu)\nonumber\\
&&+4x(PS+QR)(1-\lambda_{\bar \nu}\lambda_\nu)
+2(PQ+RS)(1+x^2)(\lambda_\nu-\lambda_{\bar \nu})\nonumber\\
&&+4x(PR+QS)(\lambda_\nu-\lambda_{\bar \nu}),\\
\gamma Z^0:{\cal F}_3&=&F_1[P(1+x^2)(1-\lambda_{\bar \nu}\lambda_\nu)
+Q(1+x^2)(\lambda_\nu-\lambda_{\bar \nu})\nonumber\\
&&+2xR(\lambda_\nu-\lambda_{\bar \nu})
+2xS(1-\lambda_{\bar \nu}\lambda_\nu)]\nonumber\\
&&+F_2[P(2-\lambda_{\bar \nu}\lambda_\nu x^2)+Q(\lambda_\nu-\lambda_{\bar\nu})(1-\frac{1}{2}x^2)\nonumber\\
&&+\frac{3}{2}xR(\lambda_\nu-\lambda_{\bar\nu})+xS(2-\lambda_{\bar\nu}\lambda_\nu)],
\end{eqnarray}

\noindent where explicitly P, Q, R, and S are

\begin{eqnarray}
P&=&[(c_{\phi}-\frac{s_{\phi}}{r_W})^{2}+\Gamma(s_\phi +\frac{c_\phi}{r_W})^2]g_V,\nonumber\\
Q&=&(c_{2\phi}-\frac{{s^{2}_{W}}}{r_{W}}s_{2\phi})(\Gamma -1)g_A,\nonumber\\
R&=&(c_{2\phi}-\frac{{s^{2}_{W}}}{r_{W}}s_{2\phi})(\Gamma -1)g_V,\\
S&=&[(c_{\phi}+r_{W}s_{\phi})^2+\Gamma(s_\phi -r_Wc_\phi)^2]g_A.\nonumber
\end{eqnarray}

\section{Formulas for the helicities}

\subsection{Formulas with right currents}

We only take the part of interference Eq. (14) for the analysis. To simplify,
we define the functions ${\cal H}_1$ and ${\cal H}_2$ from the following manner

\begin{equation}
{\cal F}=F_1{\cal H}_1(\phi,M_{Z^0_2}, x, \lambda_{\bar \nu},\lambda_{\nu})_{LRSM}
         +F_2{\cal H}_2(\phi,M_{Z^0_2}, x, \lambda_{\bar \nu},\lambda_{\nu})_{LRSM},
\end{equation}

\noindent where

\begin{eqnarray}
{\cal H}_1(\phi,M_{Z^0_2}, x, \lambda_{\bar \nu},\lambda_{\nu})_{LRSM}
&=&R_1(s)[P(1+x^2)(1-\lambda_{\bar \nu}\lambda_{\nu})+Q(1+x^2)(\lambda_{\nu}-\lambda_{\bar \nu})\nonumber\\
          && +2xR(\lambda_{\nu}-\lambda_{\bar \nu})+2xS(1-\lambda_{\bar \nu}\lambda_{\nu})],
\end{eqnarray}

\begin{eqnarray}
{\cal H}_2(\phi,M_{Z^0_2}, x, \lambda_{\bar \nu},\lambda_{\nu})_{LRSM}
&=&R_1(s)[P(2-\lambda_{\bar \nu}\lambda_{\nu}x^2)+Q(1-\frac{x^2}{2})(\lambda_{\nu}-\lambda_{\bar \nu})\nonumber\\
          && +\frac{3}{2}xR(\lambda_{\nu}-\lambda_{\bar \nu})+xS(2-\lambda_{\bar \nu}\lambda_{\nu})],
\end{eqnarray}

\noindent these functions depend on the parameters of the LRSM, on the helicities
of the neutrino and on the scattering angle.

From Eqs. (22) and (23) we consider four combinations of helicities for the neutrino
(antineutrino), obtaining the following:

Case 1

Neutrino and antineutrino with positive helicity

\begin{eqnarray}
{\cal H}_1(\phi,M_{Z^0_2}, x, \lambda_{\bar \nu}=\lambda_{\nu}=1)_{LRSM}&=&0,\nonumber\\
{\cal H}_2(\phi,M_{Z^0_2}, x, \lambda_{\bar \nu}=\lambda_{\nu}=1)_{LRSM}&=&R_1(s)[P(2-x^2)+Sx].
\end{eqnarray}

Case 2

Neutrino and antineutrino with negative helicity

\begin{eqnarray}
{\cal H}_1(\phi,M_{Z^0_2}, x, \lambda_{\bar \nu}=\lambda_{\nu}=-1)_{LRSM}&=&0,\nonumber\\
{\cal H}_2(\phi,M_{Z^0_2}, x, \lambda_{\bar \nu}=\lambda_{\nu}=-1)_{LRSM}&=&R_1(s)[P(2-x^2)+Sx].
\end{eqnarray}

Case 3

Neutrino with positive helicity and antineutrino with negative helicity

\begin{eqnarray}
{\cal H}_1(\phi,M_{Z^0_2}, x,\lambda_{\bar \nu}=-1, \lambda_{\nu}=1)_{LRSM}
&=&R_1(s)[2(P+Q)(1+x^2)+4x(R+S)],\nonumber\\
{\cal H}_2(\phi,M_{Z^0_2}, x, \lambda_{\bar \nu}=-1, \lambda_{\nu}=1)_{LRSM}
&=&R_1(s)[P(2+x^2)+Q(2-x^2)+3x(R+S)].
\end{eqnarray}

Case 4

Neutrino with negative helicity and antineutrino with positive helicity

\begin{eqnarray}
{\cal H}_1(\phi,M_{Z^0_2}, x, \lambda_{\bar \nu}=1, \lambda_{\nu}=-1)_{LRSM}
&=&R_1(s)[2(P-Q)(1+x^2)-4x(R-S)],\nonumber\\
{\cal H}_2(\phi,M_{Z^0_2}, x, \lambda_{\bar \nu}=1, \lambda_{\nu}=-1)_{LRSM}
&=&R_1(s)[P(2+x^2)-Q(2-x^2)-3x(R-S)].
\end{eqnarray}

\subsection{Formulas without right currents}

In this case we calculate the functions ${\cal H}_1$ and ${\cal H}_2$ in the
absence of right currents. This is obtained taking the limit when the mixing
angle $\phi =0$ and $M_{Z^0_2}\to \infty$ so that $\Gamma \to 0$.  In this
limit $P=g_V$, $Q=-g_A$, $R=-g_V$, $S=g_A$ and the functions ${\cal H}_1$ and
${\cal H}_2$ Eqs. (22) and (23) take the form

\begin{eqnarray}
{\cal H}_1(x, \lambda_{\bar \nu},\lambda_{\nu})
&=&R_1(s)[\{g_V(1+x^2)+2xg_A\}(1-\lambda_{\bar \nu}\lambda_\nu)\nonumber\\
           && -\{g_A(1+x^2)+2xg_V\}(\lambda_{\nu}-\lambda_{\bar \nu})],\\
{\cal H}_2(x, \lambda_{\bar \nu},\lambda_{\nu})
&=&R_1(s)[g_V(2-\lambda_{\bar \nu}\lambda_\nu x^2)-g_A(1-\frac{x^2}{2})(\lambda_\nu-\lambda_{\bar \nu})\nonumber\\
           && -\frac{3}{2}g_Vx(\lambda_\nu-\lambda_{\bar \nu})+g_Ax(2-\lambda_\nu \lambda_{\bar \nu})].
\end{eqnarray}

In a similar manner as in the previous section, we consider the following
states of helicity of the neutrino (antineutrino):

Case 1

Neutrino and antineutrino with positive helicity

\begin{eqnarray}
{\cal H}_1(x, \lambda_{\bar \nu}=\lambda_{\nu}=1)&=&0,\nonumber\\
{\cal H}_2(x, \lambda_{\bar \nu}=\lambda_\nu =1)&=&R_1(s)[g_V(2-x^2)+g_Ax].
\end{eqnarray}

Case 2

Neutrino and antineutrino with negative helicity

\begin{eqnarray}
{\cal H}_1(x, \lambda_{\bar \nu}=\lambda_{\nu}=-1)&=&0,\nonumber\\
{\cal H}_2(x, \lambda_{\bar \nu}=\lambda_\nu =-1)&=&R_1(s)[g_V(2-x^2)+g_Ax].
\end{eqnarray}

Case 3

Neutrino with positive helicity and antineutrino with negative helicity

\begin{eqnarray}
{\cal H}_1(x, \lambda_{\bar \nu}=-1, \lambda_{\nu}=1)
&=&2R_1(s)(g_V-g_A)(1-x)^2,\nonumber\\
{\cal H}_2(x, \lambda_{\bar \nu}=-1, \lambda_\nu =1)&=&R_1(s)[g_V(2+x^2)-g_A(2-x^2)+3x(-g_V+g_A)].
\end{eqnarray}

Case 4

Neutrino with negative helicity and antineutrino with positive helicity

\begin{eqnarray}
{\cal H}_1(x, \lambda_{\bar \nu}=1, \lambda_{\nu}=-1)&=&2R_1(s)(g_V+g_A)(1+x)^2,\nonumber\\
{\cal H}_2(x, \lambda_{\bar \nu}=1, \lambda_\nu =-1)&=&R_1(s)[g_V(2+x^2)+g_A(2-x^2)+3x(g_V+g_A)].
\end{eqnarray}

In the following section we analyze the angular distribution of the neutrino (antineutrino),
and interpret the cases obtained for the four combinations of helicity.

\section{Results}

The experiments of collision in the accelerators give results that depend on the
collision energy $E$ between the electron and the positron. We consider
energies available in the actual accelerators, that is, $\sqrt{s}=100$ $GeV$ \cite{Review}.
This energy is distributed between the particles that collide; then the center-of
mass energy varies by a few $GeV$ and up to $E=50$ $GeV$. The mass of the $Z^0_1$
is $M_{Z^0_1}=91.2$ $GeV$ \cite{Review}, therefore resonance exists when $E=45.6$ $GeV$,
that is, when $\sqrt{s}=2E=91.2$ $GeV$. This is manifest in the factor of resonance
$R_1(s)$, Eq. (16).

We first analyze the different states of helicity of the neutrino (antineutrino)
and subsequently the angular distribution of the pair production of neutrinos.

In the standard model, the neutrino has negative helicity $(\lambda_\nu=-1)$,
and the antineutrino has positive helicity $(\lambda_{\bar \nu}=1)$. Immediately
we interpret the cases obtained for the four combinations of helicity, with
and without right currents, Eqs. (24)-(27) and (30)-(33).

From the Eqs. (24) and (30), we observe that the antineutrino appears with
normal helicity while the neutrino is created with the opposite helicity.
It is clear from these equations that the magnetic moment induces change of
helicity to have right currents or not.

In this case, Eqs. (25) and (31) the antineutrino has helicity opposite to
normal, while the neutrino has normal helicity. The magnetic moment induces
a change in the helicity, independently of the right-handed currents.

Now, both the neutrino and the antineutrino are created with the helicities
opposite to normal, Eqs. (26) and (32). In this case, both functions
${\cal H}_1$ and ${\cal H}_2$ contribute.

The neutrino and the antineutrino appear with the normal helicities, Eqs. (27)
and (33). This corresponds to the standard model extended to the case of neutrinos
with electromagnetic interaction. This situation has already been calculated in the
collision $\nu e\to \nu e$ without right currents \cite{Vogel} and measurement
experimentally \cite{CHARM II}.

The numerical computation of the functions ${\cal H}_1$, and ${\cal H}_2$ with
and without right currents is present in the Figs. 2-7. According to the experimental
data, the allowed range for the mixing angle between $Z^0_1$ and $Z^0_2$ is
$-0.009 \leq \phi \leq 0.004$ with a 90 $\%$ C.L. \cite{J.Polak1,L3,Maya}. We
chose $M_{Z^0_2}= 500$ $GeV$ \cite{Review}. This figure does not take into account
the values of the charge radius $\langle r^2 \rangle$, and the magnetic moment
$\mu_\nu$; therefore the analysis is independent from the manner in which we
derive these quantities.

Fig. 2, shows ${\cal H}_1(\phi,M_{Z^0_2}, \lambda_{\bar \nu}, \lambda_{\nu}, x)_{LRSM}$
for the four states of helicities of the neutrino, and as function of the
scattering angle $x=\cos\theta$, with $\phi = -0.005$, and $M_{Z^0_2}= 500$ $GeV$.
We consider the energy $E= 40$ $GeV$, that is to say, $\sqrt{s}=80$ $GeV$ before
the resonance of the $Z^0_1$. The unities in the vertical scale are arbitrary.
We observed that ${\cal H}_1(-1, 1)_{LRSM}$ have an increasing behavior, while
${\cal H}_1(1, -1)_{LRSM}$ have a decreasing behavior. In this case ${\cal H}_1(\pm 1, \pm 1)_{LRSM}=0$.

In Fig. 3, again graphic the functions ${\cal H}_1(1,-1)_{LRSM}$ and ${\cal H}_1(-1,1)_{LRSM}$
with the same data, except that now the energy of collision is $E=50$ $GeV$,
that is to say, for resonance above $Z^0_1$. Again the functions have an 
increasing and decreasing behavior, only that now these change sign.

Fig. 4 shows ${\cal H}_2(\phi,M_{Z^0_2}, \lambda_{\bar \nu}, \lambda_{\nu}, x)_{LRSM}$,
with $\phi =-0.005$, $M_{Z^0_2}=500$ $GeV$ and $E=40$ $GeV$. The functions
have an increasing and decreasing behavior however, ${\cal H}_2(-1, 1)_{LRSM}$ is increasing,
while ${\cal H}_2(1, -1)_{LRSM}$ is decreasing, and ${\cal H}_2(1, 1)_{LRSM}={\cal H}_2(-1, -1)_{LRSM}$.

Fig. 5 shows again ${\cal H}_2(\phi,M_{Z^0_2}, \lambda_{\bar \nu}, \lambda_{\nu}, x)_{LRSM}$
with the same data as $\phi$ and $M_{Z^0_2}$ in Fig. 4, only now with
$E=50$ $GeV$. The behavior is similar except for a change of sign proceeding of
the sign relative to $s$ and $M^2_{Z^0_2}$ in $R_1(s)$ Eq. (16).

Finally, in Figs. 6, 7 we show the functions ${\cal H}_1$ and ${\cal H}_2$
for $\phi=-0.005$, $M_{Z^0_2}=500$ $GeV$, and $E=40$ $GeV$, with and without
the contribution of the parameters of the  LRSM. We observe than the contribution
of the right-handed currents is small.

\section{Conclusions}

In this paper, we have calculated the total cross-section of the pair production
of neutrinos and we also analyze the differents states of helicity of the neutrino
(antineutrino), as well as the angular distribution of the neutrino (antineutrino)
with respect to the original direction of the electron (positron) to different
states of helicity of the neutrino.

We find than the favored directions of the neutrino (antineutrino) with respect
to the electron (positron) direction are forward ($\theta =0$) and backward ($\theta = \pi$)
and is not very probable in the perpendicular direction ($\theta = \frac{\pi}{2}$).

The angular distributions that before the resonance are constructive or
destructive, after resonance inverted their character. The angular distributions
are more sensitive to the changes of helicity than to the contributions of the
right-handed currents.

The existence of the magnetic moment favors the creation of pairs with one of
the two neutrinos with the helicity opposite to the normal.

Only the right-handed currents favors the creation of neutrinos and antineutrinos
both with helicities opposite to normal.

\vspace{1cm}

\begin{center}
{\bf Acknowledgments}
\end{center}
Financial support from {\it Consejo Nacional de Ciencia y Tecnolog\'{\i}a}
(CONACyT), the {\it Sistema Nacional de Investigadores} (SNI) M\'exico
and the {\it Programa de Mejoramiento al Profesorado} (PROMEP) is gratefully acknowledged.
The authors would also like to thank Anna Maria D'Amore for revising the manuscript.

\newpage

\begin{center}
FIGURE CAPTIONS
\end{center}

\bigskip

\noindent {\bf Fig. 1} The Feynman diagrams contributing to the process
          $e^-e^+\to \nu \bar \nu $, in a left-right symmetric model.

\bigskip

\noindent {\bf Fig. 2} Plot of ${\cal H}_1(\phi,M_{Z^0_2}, \lambda_{\bar \nu}, \lambda_{\nu}, x)$
as a function of the scattering angle $x=\cos\theta$ with $\phi =-0.005$,
$M_{Z^0_2}=500$ $GeV$, and $\sqrt{s}=80$ $GeV$.

\bigskip

\noindent {\bf Fig. 3} Same as in Fig. 2, but with $\sqrt{s}=100$ $GeV$.

\bigskip

\noindent {\bf Fig. 4} Plot of ${\cal H}_2(\phi,M_{Z^0_2}, \lambda_{\bar \nu},\lambda_{\nu}, x)$
as a function of the scattering angle $x=\cos\theta$ with $\phi =-0.005$,
$M_{Z^0_2}=500$ $GeV$, and $\sqrt{s}=80$ $GeV$.

\bigskip

\noindent {\bf Fig. 5} Same as in Fig. 4, but with $\sqrt{s}=100$ $GeV$.

\bigskip

\noindent {\bf Fig. 6} Plot of ${\cal H}_1$ for $\sqrt{s}=80$ $GeV$ with and
without the contributions of the parameters of the LRSM.

\bigskip

\noindent {\bf Fig. 7} Same as in Fig. 6, but for ${\cal H}_2$.

\end{document}